\documentclass[a4paper, 11pt]{article}
 
\usepackage[sc]{mathpazo}
\usepackage{chicago}
\usepackage{pifont}
\usepackage{graphicx}
\usepackage{endnotes}
\usepackage{authblk}                                           
\usepackage{amssymb}
\usepackage{bbm}    
\usepackage{pgf}
\usepackage{tikz}
\usetikzlibrary{positioning,shapes,arrows,automata}
\usepackage{tensor}                                            
\usepackage{amsmath}

\usepackage{setspace}                           
\usepackage{hyperref}
\usepackage{epigraph}
\usepackage[utf8x]{inputenc}
\usepackage[left=1in,right=1in,top=1in,bottom=1in]{geometry}                                  
\usepackage{enumitem}

\usepackage{multirow}

\hypersetup{                 
    colorlinks=true,         
    linkcolor=blue,          
    citecolor=blue,           
    urlcolor=blue          
}
\let\oldmarginpar\marginpar
\renewcommand\marginpar[1]{\oldmarginpar{\color{red}\raggedright\scriptsize #1}}

\newcommand{\comment}[1]{}

\def\t{\ensuremath\mathrm}

\title{{\bf On the Empirical Consequences of the AdS/CFT Duality}}

\author[1]{\bf Radin Dardashti  \thanks{email: \href{mailto:dardashti@uni-wuppertal.de}{dardashti@uni-wuppertal.de}}}

\author[2]{\bf Richard Dawid \thanks{email: \href{mailto:richard.dawid@philosophy.su.se}{richard.dawid@philosophy.su.se}}}

\author[3,4]{\bf Sean Gryb\thanks{email: \href{mailto:sean.gryb@gmail.com}{sean.gryb@gmail.com}}}

\author[3]{\bf Karim Th\'ebault\thanks{email: \href{mailto:karim.thebault@bristol.ac.uk}{karim.thebault@bristol.ac.uk}}}

\affil[1]{\small{{\it Department of Philosophy}, University of Wuppertal}} 

\affil[2]{\small{{\it Department of Philosophy}, University of Stockholm}}

\affil[3]{\small{{\it Department of Philosophy}, University of Bristol}} 

\affil[4]{\small{{\it H. H. Wills Physics Laboratory}, University of Bristol}}

\begin{document}

\date{}
\maketitle

\begin{abstract}
We provide an analysis of the empirical consequences of the AdS/CFT duality with reference to the application of the duality in a fundamental theory, effective theory and instrumental context. Analysis of the first two contexts is intended to serve as a guide to the potential empirical and ontological status of gauge/gravity dualities as descriptions of actual physics at the Planck scale. The third context is directly connected to the use of AdS/CFT to describe real quark-gluon plasmas. In the latter context, we find that neither of the two duals are confirmed by the empirical data.

\end{abstract}

\section{Introduction}

The mathematical, physical and conceptual consequence of dualities has over the last decade been subject to growing philosophical literature.\footnote{See for example \cite{dawid:2007,rickles:2011,dawid:2013,matsubara:2013,Haro:2015,Fraser:2017,Rickles:2017,Haro:2017a,Haro:2017,Huggett:2017}.} There has not, however, hitherto been any detailed philosophical analysis of the \textit{empirical} consequences of dualities. This paper aims to address this deficit by considering a cluster of questions relating to the empirical consequences of the AdS/CFT duality. Before we outline the goals of our discussion it will be instructive to review the details of the duality itself.\footnote{Here and below we are, for the most part, following the excellent treatment of \cite{ammon:2015}. For historical details on AdS/CFT duality see \cite[\S10.2]{rickles:2014}.}  The AdS/CFT duality relates the physics of `bulk' gravity theories on asymptotically anti-de Sitter spacetimes to that of `boundary' conformal field theories. While there are many concrete examples of this duality\footnote{See e.g. \cite{horowitz2009gauge} and \cite{kaplan2013lectures}.}, we will be focusing  on the most discussed example of the AdS/CFT duality, namely between $SU(N)$, $\mathcal{N}=4$ Super Yang-Mills theory in 3+1 dimensions and type IIB superstring theory on AdS$_5\times$ S$^{5}$ introduced by \cite{maldacena1999large} and further developed in   \shortcite{gubser1998gauge} and \cite{Witten:1998qj}.

The strong form of the duality is defined in the 't Hooft  limit where the number of colour charges $N\rightarrow\infty$ but the 't Hooft  coupling $\lambda:=g^2_{YM}N$ is fixed. In this context, the duality is between the expectation value of the Wilson loop operator for an SU(N), $\mathcal{N}=4$ Super Yang-Mills theory in a flat spacetime and the semi-classical partition function of a macroscopic string in AdS$_5\times$ S$^{5}$ whose worldsheet $\Sigma$ ends on the path of the Wilson loop at the boundary. This is thus a gauge/gravity duality with the Super Yang-Mills theory playing the `gauge part' and the string in AdS$_5\times$ S$^{5}$ playing the `gravity' part. The duality is  between the mathematical objects that represent all possible empirical observations on two sides. It is crucial to note that the existence of this duality as an exact relation is largely conjectural: it remains mathematically unproven despite being extensively verified in explicit cases. In what follows we will put the formal status of the duality to one side and simply assume that it holds. Furthermore, for ease of reference we will refer to this precise specification of the duality below as `AdS/CFT' with the relevant specifications implicit.  

To return to the goals of the paper. Given that our aim is to assess the empirical consequences of dualities an immediate question is why we have chosen to focus on the AdS/CFT duality. There are a number of obvious reasons why both sides this duality might be taken to be empirically irrelevant. First, anti-de Sitter spacetimes are characterised by a negative cosmological constant, so the gravity side of the duality is evidently not well-suited to describe the geometry of our universe which according to observation has a positive cosmological constant. Second, on the field theory side, we have good reasons to expect that the number of colour charges is finite. Third, none of the empirically adequate quantum field theories that make up the standard model are conformally invariant. It is clear, therefore, that according to our best current understanding our world is not AdS and none of the known interactions are described by a conformal field theory. Thus the duality is applicable neither to the world as a whole nor for a description of nuclear interactions. 

Why then consider the empirical consequences of AdS/CFT? Our two main reasons are related to two very different applications. First, we hope to learn something about our actual world by grasping the precise duality. This may happen either due to a so far undiscovered general gauge/gravity duality or, more modestly, if some aspects of AdS/CFT can be applied to other contexts as well.\footnote{For example see \cite{harlow2016jerusalem} for an application to the study of the unitarity of black hole physics} In the latter case, one hopes for \textit{conceptual} similarities between the AdS/CFT case and the so far insufficiently understood theory describing actual physics at the Planck scale. Second, in recent years AdS/CFT has been applied to contexts, like a quark-gluon plasma near criticality, which are not described by a CFT but which turn out to be quantitatively approximately describable by a CFT (which, in turn is calculable in the dual gravity theory). In that case, there \textit{are} empirical similarities between predictions of the AdS/CFT theory and the real world. Thus there are good reasons to be interested in the empirical consequences of the AdS/CFT duality, despite the apparent strangeness of the topic.  

The natural starting point for a philosophical investigation into the empirical consequences of an area of science is to consider questions relating to confirmation.  The following strike us as particularly pertinent: 1) Must we re-understand the role of inductive evidence once dualities come into play?; 2) Are both sides of a duality equivalently confirmed by collection of supporting empirical evidence?; 3) What are the implications of our view on confirmation for scientific ontology?; and 4)  What are the empirical consequences of using dualities to model quark-gluon plasma? Although its use in scientific practice is rather varied, what we mean here and below by `confirmation' is the specific term of art as used in contemporary philosophy of science. That is, confirmation as a relation of inductive support between some evidence and some hypothesis. Significantly, we will, for the most part, below always be making reference to notions of confirmation that that are not putatively conclusive: we will almost always be talking about whether or not there is \textit{partial} inductive support of a conclusion by evidence. A precise framework for analysing such relations of partial inductive support in terms of `personal probabilities' (or credences) is Bayesian confirmation theory, and we will apply this framework extensively below.\footnote{For models of confirmation in terms of the Bayesian framework see \shortcite{hartmann2010bayesian,bovens2004bayesian}.} 

Our analysis will be conducted with reference to three distinct contexts. The first context follows from the observation that, although the AdS/CFT duality is itself extremely unlikely to feature in a fundamental or effective description of our universe, it is a well-studied exemplar of the general type of gauge/gravity duality that could be expected to be physically salient. As such, we adopt the epistemic standpoint of an agent situated in a universe where AdS/CFT is a feature of the \textit{fundamental} theoretical framework representing the world. Although in this sense rather wildly hypothetical, we expect our analysis be broadly applicable to the empirical and ontological implications of any fundamental gauge/gravity duality. The second line of inquiry considers a distinct hypothetical scenario where the duality relation is a characteristic of an  \textit{effective} description of the world that does not hold at the fundamental level. Although this context is importantly different from the first, we find that many aspects of our analysis carry through. The third line of inquiry addresses the way the AdS/CFT duality has been put to empirical use in the description of quark-gluon plasmas. In such contexts the duality is combined with a limiting relation between conformal field theories and certain parameter regimes of quantum field theories like QCD. We find that in this context the empirical consequences of the AdS/CFT duality must be understood in a very different way. In particular, we conclude that one should not take the partially empirically successful application the AdS/CFT duality to the study of quark-gluon plasmas to have told us anything regarding the empirical status of string theory.
 
\section{Fundamental Theory Context}

In this section we will concern ourselves with the empirical consequences of AdS/CFT seen as part of a fundamental theory. We will adopt the epistemic standpoint of an agent situated in a universe where the AdS/CFT correspondence is part of a theory which is believed to provide a \textit{fundamental predictive framework}. A fundamental theory is a framework for producing predictions concerning phenomena in a given domain that does not provide an effective theory for, or approximation to, an underlying theory that alters at least some of its core posits. Note that, while a theory is factually either fundamental or not, fundamentality can only be claimed based on an agent's epistemic standpoint, which may change in time. Newtonian mechanics was taken to be fundamental in the given sense by most physicists in the nineteenth century. It is not taken to be fundamental today. 

Our notion of `fundamental theory' is a narrow one in two senses. First, by `fundamental' we do not mean `universal'. While fundamentality does imply that the theory remains viable to arbitrarily small distance scales, there is no assumption that a fundamental theory provides a complete description of all physical facts. For example, in the nineteenth century Newtonian mechanics could be understood as a fundamental theory but not applicable to the domain of electromagnetic phenomena. It would thus be non-universal.  Our line of inquiry here thus differs substantially from one pursued by \citeN{Haro:2017} who consider the implications of dualities in the context of `a physical theory as a putative theory of the whole universe, i.e. as a putative cosmology, so that according to the theory there are no physical facts beyond those about the system (viz. universe) it describes' (p.37-38). Theories that are universal in this sense are necessarily fundamental in our sense, but fundamental theories are not necessarily universal.\footnote{The question of whether or not universal theories are possible or in principle would involve wading into the deep and murky waters of the reduction/emergence debate. Something we are at pains to avoid.}


The second aspect of narrowness in our definition of fundamental is particularly  important. 
Although fundamentality, in our sense, does relate to non-empirical aspects of a theory it does not, on our view, render a theory entirely closed to conceptual revision. 
The precise definition of fundamentality is particularly sensitive in the context of string dualities. It is sometimes suggested that M-theory, which is conjectured to stand in an exact duality relation (and therefore to be exactly empirically equivalent) to certain types of superstring theory, is more fundamental than the superstring formulations because it offers more direct access to the degrees of freedom that govern string physics beyond the perturbative regime. On our definition of fundamentality, M-theory is not more fundamental than, let's say, Type IIA superstring theory for the very reason that the two theories are (by conjecture) empirically equivalent to each other. 

Thus, what we call a fundamental theory context can be indicated more precisely in two steps. First, a theory is delimited in terms of a set of core theoretical posits and a target domain of empirical phenomena. Second, the theory is a fundamental theory if and only if there is no further theory that is more empirically adequate in the target domain and does not share the same core theoretical posits. A theory that is `superseded' by a successor theory with the very same empirical predictions but a different set of fundamental posits, will still count as a fundamental theory.  

Now that we have some clear sense of what we mean when we talk about fundamental theories, it will be useful to attempt to achieve some clarity regarding what we mean by \textit{empirically equivalent} theories. The most well-known and discussed notion of equivalence in the context of scientific theories is empirical equivalence. A minimal schema for describing such form of equivalence can be specified as follows. Take a pair of \textit{uninterpreted} theories as being each given by some pair of linguistic specifications (models, sentences, categories, equations, words....) that we denote as $T1$ and $T2$ respectively. A minimally instrumentalist interpretation, $\mathcal{II}$, of such theories would be a mapping that connects linguistic items (or groups of items)  to observables (pointer readings, scattering amplitudes, physical quantities...) that are either introduced in terms of a further langauge or extra-linguistically. An ontological interpretation, $\mathcal{OI}$ of such theories would be a richer mapping that includes connections between linguistic items (or groups of items) and non-observables (entities, objects, structures...) such that non-observables are understood (in some sense) to be existent things represented by the linguistic items. Empirical equivalence is then given by a situation where  $\mathcal{II}(\text{T1})= \mathcal{II}(\text{T2})$. 

A sufficient, but not necessary, condition for empirical equivalence is the existence of a mapping \textit{at the linguistic level} that is such that all linguistic items that are suitable for interpretation as observables are appropriately connected. Such a `translation manual'  between theories is precisely what is found in the case of dualities in general and the AdS/CFT duality in particular. It is easy to see why AdS/CFT implies empirical equivalence as per above: given the reasonable assumption that the expectation value of the Wilson loop operator on the gauge side and full partition function on the gravity side are the \textit{only} things that minimally instrumentalist interpretation connects to observables, we immediately have that $\mathcal{II}(\text{AdS})=\mathcal{II}(\text{CFT})$ in the respective semi-classical and 't Hooft limit. The AdS/CFT duality gives two different groups of linguistic items that describe \textit{the same observables}.\footnote{Much more could be said, of course, towards a fuller characterisation of both empirical equivalence in general and its relation to a formal characterisation of dualities. Giving such an account is explicitly not our purpose here. This notwithstanding, we do take our minimal schema to be sufficient to our current purpose of making clear the empirical significance of dualities in our three contexts. For a much less minimal account of duality and equivalence see \cite{Haro:2017} which builds upon \cite{Haro:2016,Haro:2017a}. For further recent work on a similar theme see \cite{Fraser:2017,Rickles:2017}.}  

These definitions and qualifications now made, we can proceed to our principal line of analysis. We would like to adopt the epistemic standpoint of an agent situated in a universe where the AdS/CFT correspondence is a core posit of a fundamental theory. To what extent should we take confirmation of the theory by evidence to be suitably dual? That is, could an agent take only one side of the duality as supported by the empirical success of the theory or must they consider empirical evidence for one side of the duality as necessarily evidence for the other? The inferential standpoint of our hypothetical agent can be modeled in terms of Bayesian confirmation theory as follows. We denote by $\t H_1$ the proposition:
\begin{description}
\item[$ \t H_1$]: Type IIB supersting theory on AdS$_5\times S^5$ provides an empirically adequate description of the target system $\mathcal{T}$ within a certain domain of conditions $D_\text{AdS}$ that include the semi-classical limit.
\end{description}
 Similarly, we denote by $\t H_2$:
\begin{description}
\item[$ \t H_2$]: $SU(N)$, $\mathcal N = 4$ SYM in $3+1$ provides an empirically adequate description of the target system $\mathcal{T}$ within a certain domain of conditions $D_\text{CFT}$ that include the 't Hooft limit.
\end{description}
The negation of these propositions are defined accordingly. Finally, we can encode the evidence available to our agent by the propositions  $\t E$, the empirical evidence obtains, and $\neg \t E $, the empirical evidence does not obtain. 

Let us assume that the agent has derived the evidence from the AdS side of the duality. Since this is the weakly coupled side this will invariably be technically more straightforward. We thus have that $\t H_1 \rightarrow \t E$ which means (almost trivially) that $P(\t H_1|\t E)>P(\t H_1)$, given that $0< P(\t H_1) < 1$ and $0< P(\t E) < 1$, and so we have confirmation. It is instructive to break this very standard inference down a little.  By Bayes' theorem: 
\begin{equation}
P(\t H_1|\t E)=\frac{P(\t E\t | H_1)P(\t H_1)}{P(\t E)}
\end{equation}
The marginals, $P(\t H_1)$ and $P(\t E)$, we take to be $x\in\mathbb{R}$ s.t. $0<x<1$, but otherwise rationally unconstrained. Given that we have assumed that $\t H_1 \rightarrow \t E$ it follows necessarily that the likelihood is equal to one, i.e., $P(\t E\t | H_1)=1$. It then immediately follows that the confirmation measure $\triangle_1=P(\t H_1|\t E)-P(\t H_1)$ is greater than zero, and we have Bayesian confirmation of $\t H_1$ by $E$. 

Consider then what we can say about $\t H_2$ given $E$ in the fundamental theory context and with AdS/CFT assumed. Since we have by $\mathcal{II}(\text{AdS})=\mathcal{II}(\text{CFT})$ the empirical equivalence of both sides, it follows that  $\t H_2 \rightarrow \t E$ as well. So again we have that the likelihood should be set to one,  i.e., $P(\t E\t | H_2)=1$. Things are a little trickier regarding the marginal $P(\t H_2)$. In principle, even despite the fundamental theory context and assumption of AdS/CFT one might still set $P(\t H_1)\neq P(\t H_2)$ since these are subjective degrees of prior belief regarding the truth of logically distinct statements. We think in the fundamental theory context it seems intuitively clear that one \textit{should} set $P(\t H_1)=P(\t H_2)$ since there is no clear physical reason for assigning different priors in the fundamental case.\footnote{Here `should' is understood in some normatively weaker sense than `an agent is rationally compelled to'.} However, clearly the sign of the confirmation measure will be positive irrespective of the prior: so long as $0<P(\t H_2)<1$, we already have $\triangle_2>0$ and thus we have confirmation of $\t H_2$ by $E$. Moreover, if we define the relative confirmation measure as:
\begin{equation}
 \bar{\triangle}_i=\frac{P(\t H_i|\t E)-P(\t H_i)}{P(\t H_i)}
\end{equation}
then just by the equation of likelihoods we immediately have that $\bar{\triangle}_1=\bar{\triangle}_2$, irrespective of the priors for $\t H_1$ and $\t H_2$. This point will be significant for the discussion of the effective theory context below. 
 
In addition to questions regarding empirical adequacy, we might also consider the inferences regarding further ontological interpretation that our hypothetical agent might make based upon the evidence obtained. There are three possible options corresponding to the popular philosophical positions on the realist-antirealist spectrum. Most straightforwardly, an ontic structural realist \cite{Ladyman:2016} would plausibly argue for an ontological interpretation such that the common structure between the two sides of the duality is reified. A shared structural vocabulary common between the two sides of the duality that would be taken to do the work in representing the `actual structure of the world'. This would imply directly that $\mathcal{OI}(\text{AdS})=\mathcal{OI}(\text{CFT})$. Also straightforwardly, an instrumentalist \cite{stein:1989} or empiricist \cite{vanfraassen:1980} would simply refuse to license further beliefs about extra-observables and would reject any commitment to entities, objects or structures \textit{actually in the world} as faithfully being represented or not. 

The situation of a standard realist is more subtle. Such realists typically justify belief in the faithful representation of extra-observables by theoretical terms by reliance on empirical evidence combined with inference to the best explanation. Depending upon what one thinks about the explanations on offer, this might seem to allow for an ontological interpretation such that  $\mathcal{OI}(\text{AdS})\neq\mathcal{OI}(\text{CFT})$. In particular, our hypothetical realist agent might reason that the gravity side of the duality is more fundamental and that their universe \textit{really is} described by a string theory in anti-de Sitter spacetime and not in any way by a Super Yang-Mills theory. This would be to take the CFT seriously only as an uninterpreted framework and ignore its putative claims for representational capacity. Plausibly, such a position would be unstable since in interpreting the AdS side one is necessarily implicitly also interpreting at least the empirical sector of the CFT side. Whatever explanation of empirical regularities is appealed to for justifying belief in an ontology based upon the AdS side, a parallel justification is available for belief in an ontology of the CFT side. And since we are, by assumption, dealing with fundamental theories the virtue of such explanations for future generalisations cannot be used as a tie-breaker between them. Essentially, the joint claim that we have a fundamental predictive framework and that the ontological interpretation of such a framework can be based upon inference to the best explanation, force our hypothetical realist to accept a dual ontology at the pain of otherwise making an arbitrary, non-inferential ontological commitment.  In this vein, it has been argued by a wide range of philosophers of science \cite{dawid:2007,rickles:2011,matsubara:2013,Rickles:2017,Huggett:2017} that an exact duality renders the duals different formulations of one and the same theory.  One way to avoid this perspective would be to take dualities as approximate rather than exact \cite{Haro:2015}. An alternative is to stick to the understanding that dualities are exact but move to a context in which the dual theories are understood to be effective. In the following section we will consider the implications of the AdS/CFT duality in this effective theory context.

\section{Effective Theory Context}

Recall from above that we defined the fundamental theory context as a situation where there is no further theory that is more empirically adequate in the given target domain and does not share the same core theoretical posits. The contrast case that we will consider in this section is the `effective theory context' where there \textit{is} a further theory that is more empirically adequate in the given target domain and does not share the same core theoretical posits. There are two reasons why this is a particularly interesting context to consider dualities within. First, it addresses a possible status of exact dualities that, to our knowledge, has not previously been discussed in the literature. Second, it brings us one step closer to the application of dualities to quark-gluon plasma calculations, which shall be discussed in the final section.

What makes the effective context relevant to our investigation is its potential to break the principle of equal confirmation of dual theories that seems unassailable in the context of fundamental dualities. To be sure, effective theories, even when known not to be fundamental, can be strongly confirmed by empirical data within the theories' intended domains. Most empirical confirmation plays out in such contexts. However, situations can arise where the knowledge about a theory's limited realm of applicability may have a detrimental effect on the confirmation value of data that is in agreement with that theory's predictions.

To understand this point, let us first think about the case of two conceptually distinct theories that are not fully equivalent to each other but share some empirical implications. An experimentum cruxis, that is strong empirical confirmation of one theory and strong empirical refutation of the other in a regime where the theories' predictions differ from each other, would obviously imply the rejection of the refuted theory also as a description of those contexts where the two theories make the same predictions. Now, let us assume that the given theory has not been \textit{empirically disconfirmed} by data beyond the regime where the two theories are empirically indistinguishable but there are strong reasons to doubt that the theory can be made consistent or can be embedded in a consistent theory that covers a wider range of phenomena. Such considerations would clearly have an impact on assessments of the status of the given theory also in the constrained empirical regime where it is applicable and agrees with the available data. In other words, data that is consistent with that theory would not be taken to amount to strong confirmation of the theory because of conceptual problems related to the theory in a wider context. In Bayesian terms, this situation can be modeled by attributing a very low prior to the given theory due to the expected problems of embedding it in a wider conceptual context. A good example of such a situation is the case of modified Newtonian dynamics (MOND) \cite{milgrom:1983}. MOND is particularly well-adapted to reproduce observed galaxy rotation curves \cite{sofue:2001}. However, this is not considered by scientists to amount to substantial confirmation of MOND precisely because, despite several proposals, there is no completely satisfactory way to obtain the MOND phenomenology from a complete cosmological model of the universe.\footnote{Although there has been significant work on relativistic extensions of MOND, e.g. \cite{bekenstein:2004}, as well as attempts to derive it from a more fundamental theory, e.g. Verlinde's ``emergent gravity'' approach \cite{Verlinde:2016toy,Milgrom:2016huh} or Mannheim's conformal theory \cite{Mannheim:2011ds}, even the originator of the approach accepts that such generalisations `do not yet provide a satisfactory description of cosmology and structure formation' \cite{Milgrom:2014}.}  

The case of dualities between effective theories can be understood as a natural generalisation of this scenario. For the sake of clarity, we will now consider a more general abstract duality between two effective theories and return to specific considerations regarding the AdS/CFT afterwards. Let us first assume that we have an exact duality between two theories, T1 and T2. As discussed above, this duality guarantees empirical equivalence but allows, in principle, for the possibility of distinct ontological interpretations. Using the terminology of the previous section we would have that $\mathcal{II}(\text{T1})= \mathcal{II}(\text{T2})$ but $\mathcal{OI}(\text{T1})\neq \mathcal{OI}(\text{T2})$. Let us consider again the position of a strong realist who believes that the gravity side of the duality is more fundamental and that, notwithstanding the duality, their universe \textit{really is} described by one side. In the fundamental theory context we suggested that inferentialist justification that most realists give for their view is in conflict with this form of distinct ontological interpretation of dual pairs. Prima facie, this need not be the case in the effective theory context. Recall from above, a non-fundamental theory is one which could be replaced by a theory that is more empirically adequate in a given domain and does not share the same core theoretical posits. Now imagine that one takes the pair T1 and T2 to be effective theories to some more fundamental theory X in this sense. The case of particular interest is where the mismatch of core theoretical posits includes the dual structure of the effective description. In this situation it might seem reasonable for our strong realist to assert $\mathcal{OI}(\text{T1})\neq \mathcal{OI}(\text{T2})$ on the basis that $\mathcal{OI}(\text{T1})\approx\mathcal{OI}(\text{X})$ and $\mathcal{OI}(\text{T2})\not\approx\mathcal{OI}(\text{X})$. Plausibly then, this relative closeness of ontology combined with the empirical superiority of X would provide an inferentialist basis for realist commitment to T1 over T2, despite the fact that they are dual theories. The claim is that this realist commitment to T1 over T2 can translate into the understanding that T1 gets strong confirmation by data that agrees with its predictions while the empirically equivalent T2 does not. This happens if, in analogy to the MOND case, a much higher prior can be attributed to T1 based on its better alignment with the broader physical picture represented by X. In the given case, that would mean that T1 gets a higher prior than T2 because there is a fundamental theory X that retains the ontology of T1 while it seems unlikely that there is a fundamental theory that retains the ontology of T2.

A natural line of objection is to question whether matters of ontology can ever become relevant for matters of confirmation. There is at least one context in physics, however, where ontology indeed becomes relevant for the assessment of a physical hypothesis. Bohmian non-local hidden variable models have been established as a possible interpretation of non-relativistic quantum mechanics.\footnote{See \cite{Goldstein:2017} for an up-to-date discussion of Bohmian mechanics.} Making the Bohmian approach compatible with the Lorentz invariance of relativistic physics is an outstanding problem. If we assume, as most physicists do, that Lorentz invariance is a fundamental feature of relativistic physics then one could plausibly argue that a relativistic formulation of Bohmian quantum mechanics might in principal be ruled out. For our purposes it is not important whether or not those problems can be overcome in a way that upholds the Bohmian approach as an attractive interpretation of relativistic quantum mechanics. The crucial question for us is: \textit{if} a relativistic Bohmian quantum mechanics proved impossible, \textit{would} that then discredit non-relativistic Bohmian quantum mechanics as well, even though the latter works in its own regime? Most observers think it would do so. The failure of a set of ontological claims at a fundamental level is, in this case, taken to disfavour those claims also at the effective level. 
 
One might object that Bohmian quantum mechanics is an interpretation rather than a theory. But does this disqualify the example? If Bohmian quantum mechanics is understood as an interpretation rather than a theory because it is empirically equivalent to other `interpretations' of QM, then one might call dual theories `interpretations' as well as soon as one takes their ontic commitments seriously.  

One may also have the understanding that Bohmian quantum mechanics is an interpretation of quantum mechanics because it is essentially bound to a contingent ontic view on quantum mechanics. By its very nature, the Bohmian approach is predicated on a realist perspective. Its truth would imply the falsity of the ontological outlook of alternative interpretations of quantum mechanics. A failure of relativistic Bohmian mechanics would discredit the entire Bohmian approach to quantum mechanics for the very reason that the approach implies a position of ontological scientific realism: if the objects of Bohmian quantum mechanics are shown not to exist relativistically, then they are shown not to exist at all. In general terms, if one adopts a strong ontic realist view, then the degree of confidence one has in the adequacy of an effective theory is extremely sensitive to that theory's relationship with more fundamental theories.  

Applying a similar rationale to dualities in the effective theory context thus hinges on accepting an ontic commitment as an essential part of endorsing a physical theory. Clearly for the defender of a strict distinction between a scientific theory and its interpretation, commitment to a theory need not involve endorsing an associated ontology. Such non-realism would already mean that empirical confirmation is understood to be confined to theories and thus we should reject the possibility of different degrees of confirmation for different dual effective theories. Thus the question is whether string theory is consistent with any ontic commitment (that is, any positive position in the scientific realism debate) that can provide a basis for distinguishing between an ontic commitment to T1 and to T2. As already noted, in the fundamental theory context there are good reasons to assert that string dualities are at variance with such a strong form of ontic scientific realism \cite{dawid:2007,rickles:2011,matsubara:2013,Rickles:2017}. In particular, the possibility to transform a theory into its dual makes the choice between dual theories reminiscent of an (unphysical) gauge degree of freedom, which seems to block (or at least make less plausible) any attempt to single out the ontology of one of the duals as the true ontology \cite{Rickles:2017}. The understanding that all dual representations are equally valid is further strengthened by the fact that typically all dual perspectives are needed in order to acquire a satisfactory understanding of the physical content of string theory \cite{dawid:2013}. These lines of reasoning tacitly assume, however, that string dualities play out at a fundamental level. It thus seems possible for a defender of the strong form of ontic scientific realism about dualities, in which the realist commitment to T1 differs from that to T2, to insist that their position is made more plausible by the effective theory context. 

Whether or not a different degree of commitment to T1 than to T2 can be defended in that context thus depends on the specifics of the realist's commitment to X. From a minimal structural realist type position \cite{ladyman:2007}, we might give an explanation of a theory's `real' content based on its role as an effective theory. If we apply such a view to X then it is very difficult for us not to apply it equally to T1 and T2, notwithstanding the facts about perceived `closeness'. Equally, we can consider weak forms of scientific realism, such as perspectival realism \cite{massimi:2018} or consistent structure realism \cite{dawid:2013}, wherein we accept as `representations of reality'  distinct empirically equivalent scientific representations. Clearly, such weak realism offers us no basis for using X to retrospectively differentiate the `approximate truth' of the effective theories T1 and T2. An exponent of one of the weaker forms of realism would deny that approximate truth should be understood in terms of `ontic continuity' in the given context and therefore would assert that both effective theories must be equally confirmed.

Finally, we can consider something more like a `common core' perspective on scientific realism at the fundamental level. Such diachronic commonalities might be understood to trump synchronic `common core' considerations at the effective level, and therefore distinguish the ontological significance of one of the effective theories. Choosing a `divide-and-conquer' selective realism about X \cite{psillos:2005}, we might as well argue for a realist commitment to T1 over T2 due to the greater retention of theoretical terms key for explaining predictive success at the level of X. Let us imagine that there were a gauge/gravity duality at an effective level. A more fundamental theory X, however, is found to be strictly a gravity theory with a fundamental ontology similar to its effective string theory and did not have a dual field theory. In that case, an exponent of divide-and-conquer-style brand of scientific realism could argue that the truth of the fundamental theory X would render the effective gravity theory approximately true and the effective field theory false. Knowledge of the existence of X could then result in attributing a low prior to the effective field theory, in analogy to our examples of MOND and Bohmian mechanics. Data in agreement with both effective theories could then amount to strong confirmation of the gravity theory without being taken to be strong confirmation of the gauge theory. Adapting our Bayesian notation in an obvious way from above, it would then be plausible, seen from this strong realist perspective in the effective theory context, to set $P(T1)\gg P(T2)$ which means that $\triangle_{T1}\gg\triangle_{T2}$. Of course, we would still have that the relative confirmation was equivalent, so $\bar{\triangle}_{T1}=\bar{\triangle}_{T2}$.

To conclude, differentiating degrees of confirmation based on the continuity of ontic commitments from the effective to the fundamental level is an option in principle given certain forms of strong scientific realism. However, such a view is in tension with a variety of weaker forms of scientific realism and, unsurprisingly, any kind of non-realism. Moreover, such a view is also rather against the spirit of string physics, based on our current understanding of the theory. Nevertheless, we hope that in considering this option may be helpful for getting a better grasp of the range of possibilities one faces when thinking about the empirical consequences of duality relations.

\section{Instrumental Context}

The most surprising applications of the AdS/CFT duality in contemporary physics is in the context of quantum chromodynamics (QCD), the quantum field theoretic description of the strong nuclear force. That a duality which originated from string theoretic considerations has found application in the description of the phenomenology of hadronic physics is a rather beautiful irony, given string theory's own origin in attempts to find a phenomenological model of the strong interaction \shortcite{cappelli2012birth,rickles:2014}. These fascinating connections notwithstanding, one of the main messages of this paper is that one should be at pains not to over-interpret the strength of the AdS/QCD theory connection. Rather, as could be expected from the title of this section, we will argue that as things stand neither the ontological nor empirical implications of this application of the duality are compelling. Rather, as shall be detailed below, we think there good reasons to take the application of the AdS/CFT correspondence to hadronic physics as a \textit{purely instrumental} one. That is, we do not take partially empirically successful applications of AdS/QCD to have told us anything regarding the empirical status of string theory.\footnote{This is at least in tension with remarks of  \cite[viii]{rickles:2014} who remarks, en passent, that string theory proving to be an `empirical dud' is `highly unlikely as a general claim' since duality symmetries present (or originating) in string theory have to led to several `results that have experimental ramifications' including in the context of quark-gluon plasmas and  superconductivity.}

To understand how one might possibly use the AdS/CFT duality to model a special class of systems described by QCD, it will be instructive to consider the three fundamental senses in which QCD is \textit{unlike} a CFT.\footnote{Here we mostly follow the excellent treatment of \cite{ammon:2015}.} Recall that the particular example of AdS/CFT that we are considering is between Type IIB superstring theory on AdS$_5\times$ S$^{5}$ and $SU(N)$, $\mathcal{N}=4$ Super Yang-Mills theory in 3+1. First, and most obviously,  QCD is unlike  $\mathcal{N}=4$ $SU(N)$ Super Yang-Mills since  it is not supersymmetric. Second, QCD is not conformally invariant. Third, QCD has a finite colour number $N=3$. Getting around the first two differences will depend on finding a limit where Super Yang-Mills is effectively non-supersymmetric and where  QCD is effectively conformal. Given what we know from condensed matter physics about the emergence and disappearance of symmetries in different parameter regimes, the fact that such a limit can be found is not perhaps as surprising as it might seem. In particular, if we fix the CFT to be at finite temperature, but near a critical phase transition, then supersymmetry will be broken and if we choose QCD to be near the deconfined phase transition then we find that the theory becomes quasi-conformal. As we shall see later, the supersymmetry breaking is achieved by choosing a black hole in asymptotically AdS spacetime with finite but near-critical temperature on the gravity side of the duality. A physical basis to justify ignoring the final difference is rather less obvious, but can again be well-illustrated via the analogous approximation made in condensed matter physics. The CFT is presumed to be defined via a large $N$ expansion corresponding to the 't Hooft limit. That is, we take the colour number to infinity in the CFT. Now, clearly $3\not\approx\infty$, so how can we justify treating QCD as approximated by a theory in such a limit? An immediate analogy is to the thermodynamic limit in statistical mechanics, where one assumes that the particle number, $n$, is infinite although we of course know full and well that it is of the order of $10^{23}$. Making sense of the approximations and idealisations of these kind of infinite limits has been the subject of a considerable philosophical literature of the last decade.\footnote{Three particularly significant contributions are \cite{batterman:2002,butterfield:2011,norton:2012}.} In fact, \cite{bouatta:2015} make a direct comparison between the 't Hooft limit and the thermodynamic limit. The important point for our purposes is that in the context of a $1/N$ expansion, the difference between $3$ and $\infty$ at first order is small enough that subsisting one expansion with the other is a valid approximation provided one is only interested in order of magnitude estimates.\footnote{See \cite[Ch. 8]{coleman1988aspects} for an elemenatry introduction to the $1/N$ expansion in field theory.} 

One context were such estimates are of interest to physicists is in the context of quark-gluon plasmas. There is experimental evidence that such forms of matter should be understood as strongly coupled fluids. Strong coupling means that conventional perturbative techniques are unreliable. The standard alternative is then a lattice gauge field theory approach. However, for time-dependent processes, lattice methods become intractable and it is easier to appeal to the AdS/QCD correspondence to calculate physical values of the parameters of quark-gluon plasmas using string theory. A particularly vivid example is the \textit{jet-quenching parameter}, $\hat{q}$.\footnote{An earlier application of the AdS/QCD correspoondence can be found in \cite{kovtun2005viscosity}.} This is the mean transverse momentum acquired by a `hard probe' (i.e., a high momentum, narrow beam of hadronic matter) per unit distance traveled. A central field theory result is that  $\hat{q}$ can be expressed in terms of the correlator of Wilson lines along a particular contour. The gravity dual description is given in terms of a string action in a AdS-Schwarzschild \textit{black brane} spacetime with a finite but near-critical Hawking temperature.\footnote{As Hawking and Page illustrated in \cite{Hawking:1982dh}, black holes in AdS space undergo  a thermodynamic phase transition when the black hole size is of order of the characteristic radius of the AdS space.} In a landmark 2006 paper \cite{liu:2006} these techniques lead to a calculated value for $\hat{q}$ of 4.5 GeV$^{2}$/fm which is close to the experimental range from Brookhaven Relativistic Heavy Ion Collider of 5-15 GeV$^{2}$/fm. At the time this was the best theoretical estimate of the parameter. How should we understand the implications of this result for AdS, CFT or, in fact, QCD? This question turns out to be extremely subtle, and the remainder of this paper will be devoted to developing a line of analysis of this new `instrumental context' of dualities that is consistent with the conclusions made regarding the previous two contexts. 

A reasonable first step is to formulate our inferential standpoint in terms of Bayesian confirmation theory as per above. 
We denote by $\t H_3$ the proposition 
\begin{description}
\item[$ \t H_3$]: QCD provides an empirically adequate description of the target system $\mathcal{T'}$ within a certain domain of conditions $D_\text{QCD}$ that include the assumption of near-criticality.
\end{description}
We can encode the evidence available to our agent by introducing a variable $E'$  corresponding to the two values, $\t E'$, the empirical evidence regarding the target system $\mathcal{T'}$ obtains, and $\neg \t E' $, the empirical evidence does not obtain. In the case of the jet-quenching parameter the target system $\mathcal{T}'$ would be a quark-gluon plasma and the empirical evidence would be the \textit{measured value} of $\hat{q}$.  

In order to consider the relevance of this empirical evidence to an agent's beliefs regarding the empirical adequacy of CFT and string theory as descriptions of $\mathcal{T'}$ one can introduce two  propositions, $\t H_4$ and $ \t H_5$:
\begin{description}
\item[$ \t H_4$]: $SU(N)$, $\mathcal N = 4$ SYM in $3+1$ provides an empirically adequate description of the target system $\mathcal{T'}$ within a certain domain of conditions $D'_\text{CFT}$ that include the finite but near-critical temperature assumption.
\end{description}
\begin{description}
\item[$ \t H_5$]: Type IIB superstring theory on AdS$_5\times S^5$ provides an empirically adequate description of the target system $\mathcal{T'}$ within a certain domain of conditions $D'_\text{AdS}$ that include the finite but near-critical temperature assumption.
\end{description}
It is of course significant to note that in the definitions above we are considering the empirical adequacy of AdS/CFT as a description of a quark-gluon plasma. There are a variety of obvious senses in which both sides of the duality immediately fail to provide such descriptions. This notwithstanding, with regard to the jet-quenching parameter taken in isolation, empirical adequacy of the dual theories is worth considering. In particular,  we can expect that an agent can derive empirical predictions regarding $\mathcal{T'}$ using the calculational techniques available from the AdS side of the duality. Since this is the weakly coupled side this will invariably be technically more straightforward. In fact, as noted above, at least in 2006, this was the best available technique.  

We can represent this prediction of a value of the jet-quenching parameter that is derived via the AdS model in terms of a further binary variable $F$. We take this variable to have the value $\t F$ when the \textit{predicted value} obtains in the actual quark-gluon plasma. Thus we have that $\t H_4\rightarrow \t F$. The inferential structure of our current context is unlike that of the fundamental or effective theory context. In particular, there is no empirical regime in the physics of the quark-gluon plasma where the predictions from AdS actually match the data: $\t F \neq \t E'$ in $D_\text{QCD}$ since the predicted value for $\hat {q}$ was 4.5 GeV$^{2}$/fm which is outside the experimental range of 5-15 GeV$^{2}$/fm. This means that $E'$ actually disconfirms AdS: $P( \t H_5|E')<P( \t H_5)$. Due to the (assumed) exact duality between AdS and CFT, the latter gets disconfirmed by $E'$ as well. So we have $P(H_4|E')<P(H_4)$. 

As regards QCD as things have been set up, the relevance of $E'$ for $H_3$ is not clear. In particular, since there is only a crude approximating relation with the dual theories it is not clear, as we have formulated things, whether we should take the new evidence as positive, negative or neutral with regards to the empirically adequacy of QCD. The limiting relationship between QCD and AdS gives us a means to derive an order of magnitude approximation regarding the phenomenology of the former, using calculational techniques based upon the latter. Yet, AdS is not an effective theory of QCD phenomenology and cannot be confirmed even in the weak sense of increasing the degree of belief in  the theory's empirical adequacy in a specific regime of testing. In fact, as already noted, there are simple and obvious senses in which the AdS/CFT duals fail as empirically adequate  descriptions of a quark-gluon plasma: not least with regard to boundary conditions on the plasma and the finiteness of the colour number. Thus there is a strong sense in which \textit{any} empirical data collected from measurements on an actual quark-gluon plasma will straightforwardly contradict the empirical adequacy, considered in general terms, of AdS or CFT as a model of this system.\footnote{This is the direct counterpart to incisive remarks made by \cite[p. 1072]{butterfield:2011} in the context of the thermodynamical limit. In that context, Butterfield notes that taking the particle number to be extremely large corresponds to cosmic lengths and thus immediately makes the model `utterly unrealistic', and thus fundamentally empirically inadequate when considered in general. \cite{norton:2012} makes a similar point in the context of his distinction between approximation and idealisation.}  

One might, however, still plausibly expect that the data $\t E'$ to give us confidence in the reliability of the package of approximation techniques and theories that are being used. This is because, as noted in the original paper \cite[p.4]{liu:2006}, there are good theoretical reasons to expect that the value calculated using AdS would be a slight underestimate. There were good reasons to expect the right order of magnitude but also good reasons to expect some discrepancy. To be an accurate formalisation of the inferential standpoint of the relevant scientists we surely must incorporate this kind of background contextual knowledge regarding the models that are being employed into our Bayesian framework. 

Let us assume that there is some background contextual knowledge that when combined with the actual AdS calculation implies that the predicted value of the jet-quenching parameter should roughly approximate the measured data but not precisely coincide with it. Formally, we can introduce a proposition: 
\begin{description}
\item[$ \t C$]: Type IIB superstring theory on AdS$_5\times S^5$ gives correct to  order of magnitude empirical approximation to QCD as a description of the target system $\mathcal{T'}$ within certain domains of conditions $D'_\text{AdS}$ and $D_\text{QCD}$ respectively. 
\end{description}
Second, denote by $\delta E$ a proposition:
 \begin{description}
\item[$ \delta \t E$]: The difference between the predicted value derived via Type IIB superstring theory on AdS$_5\times S^5$ and the experimental result is correct to order of magnitude. That is $\t F \approx \t E'$ in  $D_\text{QCD}$. 
\end{description}
We then have that $(H_3 \wedge \t C) \rightarrow  \delta \t E$ and thus that $P(\t H_3 \wedge \t C|\delta \t E) > P(\t H_3 \wedge \t C)$. This is confirmation of the \textit{conjunction} of the empirical adequacy of QCD and the approximation relation between Type IIB superstring theory on AdS$_5\times S^5$ and QCD. Ideally one would then like to make a definite statement regarding the change in probabilities of each of the conjuncts. In particular, one would ideally like to claim in these circumstances that QCD is individually confirmed. However, it can be shown that the confirmation of a conjunction by some evidence is compatible with the disconfirmation of either (or both) the conjuncts considered individually \cite{atkinson:2009}. Thus, the formal model  only allows us to establish confirmation in the package of the empirical adequacy of QCD and the empirical approximation of QCD by string theory when taken together.\footnote{Informally, one plausibly \textit{would} still expect to have at least limited confirmation of $H_3$ by $\delta \t E$. Establishing the additional conditions necessary to show this formally would take us outside the remit of the current work however.}  

As a final remark we should note the relevance of recent theoretical work in calculating the jet-quenching parameter using Lattice QCD techniques \cite{panero:2014}. This approach leads to a calculated value of around 6 GeV$^{2}$/fm that is of course more consistent with the experimental range. Furthermore, that the value calculated via a more realistic model is indeed slightly above that calculated via the string theory model provides retrospective support to the original theoretical arguments for the string theory calculation being a \textit{slight} underestimate.    

\section{Conclusion}

Dualities are typically viewed in terms of exact correspondences between fundamental theories of the world. Dual models can and do occur, however, outside such fundamental theory contexts. If so, both the ontic implications and the specifics of the empirical relevance of dualities change.  We have argued in this paper that there is a fairly rigid principle of equal confirmation in the case of dualities between fundamental theories. Although this principle could be broken by assuming different priors for dual theories, one finds no physical argument for doing so. To the contrary, a number of physical arguments suggest treating duals as different perspectives on the same theory, which would rule out the attribution of different priors. This situation changes once one considers a duality relation at the level of effective theories that has no correspondence at the fundamental level. Such a scenario could, given a strong form of ontic realism, be viewed as providing a physical basis for viewing the duals as different theories that might even have different plausibility as viable effective theories about the world. Though such a point of view has its problems, it should not be discarded out of hand. Once one looks at the instrumental application of dualities in contexts where the empirically viable theory is not understood to have a dual, such as the case of quark-gluon plasma calculations, the role of theory confirmation does not depend on distinguishing between confirmation values for the individual duals. Rather, neither of the two duals gets confirmed by the data. Rough agreement of string theory calculations with data can only generate confirmation for the approximation together with the theory it is taken to approximate, which is QCD. 

\section{Acknowledgements}

We are very grateful to Nick Huggett and an anonymous referee for helpful comments on a draft manuscript and to audience members in Exeter for valuable feedback. KT and SG were supported by the Arts and Humanities Research Council (Grant Ref. AH/P004415/1). 

\bibliographystyle{chicago}
\bibliography{dual}

\end{document}